\documentclass[english,aps,prb,showpacs,superscriptaddress]{revtex4}

\usepackage{ae,aecompl}
\usepackage[T1]{fontenc}
\usepackage{textcomp}
\usepackage{amstext}
\usepackage{amssymb}
\usepackage{graphicx}
\PassOptionsToPackage{normalem}{ulem}
\usepackage{ulem}

\usepackage{babel}
\begin{document}

\title{Bandwidth-driven nature of the pressure-induced metal state of LaMnO$_{3}$}

\author{Aline Y. Ramos}

\email{aline.ramos@grenoble.cnrs.fr}

\affiliation{Institut N\'eel, CNRS et Universit\'e Joseph Fourier, BP 166, F-38042
Grenoble Cedex 9, France}

\author{Narcizo M. Souza-Neto}

\affiliation{Laborat\'orio Nacional de Luz S\'incrotron - P.O. Box 6192, 13084-971,
Campinas, Sao Paulo, Brazil}

\author{H\'elio C. N. Tolentino}

\affiliation{Institut N\'eel, CNRS et Universit\'e Joseph Fourier, BP 166, F-38042
Grenoble Cedex 9, France}

\author{Oana Bunau}

\affiliation{Institut N\'eel, CNRS et Universit\'e Joseph Fourier, BP 166, F-38042
Grenoble Cedex 9, France}

\author{Yves Joly}

\affiliation{Institut N\'eel, CNRS et Universit\'e Joseph Fourier, BP 166, F-38042
Grenoble Cedex 9, France}

\author{St\'ephane Grenier}

\affiliation{Institut N\'eel, CNRS et Universit\'e Joseph Fourier, BP 166, F-38042
Grenoble Cedex 9, France}

\author{Jean-Paul Iti\'e}

\affiliation{Synchrotron SOLEIL, L'Orme des Merisiers, Saint-Aubin, BP 48, 91192 Gif-sur-Yvette Cedex, France}

\author{Anne-Marie Flank}

\affiliation{Synchrotron SOLEIL, L'Orme des Merisiers, Saint-Aubin, BP 48, 91192 Gif-sur-Yvette Cedex, France}

\author{Pierre Lagarde}

\affiliation{Synchrotron SOLEIL, L'Orme des Merisiers, Saint-Aubin, BP 48, 91192 Gif-sur-Yvette Cedex, France}

\author{ Alberto Caneiro}

\affiliation{Centro Atomico Bariloche, CNEA e Universidad Nacional de Cuyo, 8400 S.C. de Bariloche, Argentine}

\abstract{Using X-ray absorption spectroscopy (XAS), we studied the local structure
in LaMnO$_{3}$ under applied pressure across and well above the insulator
to metal (IM) transition. A hysteretic behavior points to the coexistence
of two phases within a large pressure range (7 to 25 GPa).
The ambient phase with highly Jahn-Teller (JT) distorted $MnO_{6}$
octahedra is progressively substituted by a new phase with less-distorted
$JT$ ~ $MnO_{6}$ units. }The electronic delocalization leading
to the IM transition is finger-printed from the pre-edge XAS structure
around 30~GPa. We observed that the phase transition takes place
without any significant reduction of the JT distortion. This entails
band-overlap as the driving mechanism of the IM transition.}


\maketitle
LaMnO$_{3}$~is the parent compound of a family of doped manganates
that exhibit a multitude of electronic phases and unusual properties
with strong potential for new electronic devices\cite{Ahn-Science03}.
The compound crystallizes in the orthorhombically distorted perovskite
structure (space group Pbmn), in which every $Mn^{3+}$ ion with high-spin
configuration $t_{2g}^{3}e_{g}^{1}$ is surrounded by an octahedron
of six oxygen ligands. Under ambient conditions, the Jahn-Teller ({$JT$})
instability of the singly occupied $e_{g}$~orbitals gives rise to
cooperative distortions of the $MnO_{6}$ octahedra, which induce
orbital ordering and may be responsible for the insulating behavior
of LaMnO$_{3}$. Although the ground-state of LaMnO$_{3}$ can be
explained both by cooperative $JT$ distortion
and orbital exchange interaction, their relative importance lead to
different physics for the doped systems and is then an important issue
for predicting and optimizing physical properties. However, the separation
of the two effects is a very delicate issue~\cite{Millis-PRB96,Tyer-EPL04,Zenia-NewJPhys05,WeiGuoYin-PRL06,Yamasaki-PRL06,Fuhr-PRL08,Pavarini-PRL10}
that calls for accurate experimental probes. 

At ambient pressure LaMnO$_{3}$ undergoes at $T_{IM}$=$710\, K$
an insulator to metal (IM) transition, structurally described as a
transition from an ordered toward a disordered array of JT distorted
octahedra. The long range structure is characterized by a strong cell
symmetrization and the loss of the orbital order\cite{Rodriguez-Carvajal-PRB98}.
The local scale JT splitting persists essentially unaltered across
the IM transition, which is marked by the symmetrization of the thermal
fluctuations in the distorted $MnO_{6}$ units\cite{Araya-Rodriguez-JMMM01,Souza-PRB04}. 

At ambient temperature, transition towards a metal state can also
be attained by application of external hydrostatic pressure.
High pressure reduces lattice parameters, favoring orbital overlap
and concomitant enhancement of electron delocalization. It also forces
an atomic rearrangement, tending to reduce lattice distortions (inter-octahedral
and intra-octahedra rearrangements). In LaMnO$_{3}$
the IM transition has been reported by Loa and coworkers at an applied
pressure of 32~GPa \cite{Loa-PRL01}. Using X-ray diffraction under
pressure, these authors~obtained refined atomic positions up to 11~GPa\cite{Loa-PRL01}.
From an extrapolation of their results, they predict that the local
$JT$ distortion should completely vanish around 18~GPa. However,
in a previous X-ray absorption spectroscopy (XAS) study under pressures
up to 15~GPa~\cite{Ramos-PRB07}, our group observed a shortening
of the long MnO bond lenght too limited to lead to a quenching of
the $JT$ distortion at 18~GPa, contradicting those predictions.
Also by an extrapolation, it was deduced that a total removal of the
$JT$ distortion would occur only for pressures around 30~GPa, close
to the onset of metallization. Importantly, when the pressure was
released from 15~GPa down to 9~GPa, an incomplete recovering of
the structural environement aroused suspicion of a possible phase
coexistence in this pressure range. Phase coexistence are not unusual
in perovskite material. It has been observed in LaMnO$_{3}$ at high
temperature\cite{Prado-JMMM99,Qiu-PRL05} and even suggested at high
pressures \cite{Loa-PRL01,Trimarchi-PRB05,Ramos-JPCS09}. Recently,
Baldini and coworkers\cite{Baldini-PhysRevLett2011} have pointed
out the persistence of the $JT$ distortion coexisting with a new
emergent undistorted phase up to 32~GPa. Besides its importance for
the thermodynamics of the transition, a phase coexistence may completely
invalidate any extrapolation of the experimental data at higher pressures
and revoke previous conclusions about the vanishing of $JT$ distortion
in the metallic phase. 

In this letter we investigate by XAS the local structure
in LaMnO$_{3}$ under applied pressures across and well-above the
IM transition. XAS is a local probe, mostly affected
by the short- and medium-range environment of a selected atom and
sensing short time scale. There have been various XAS works shedding
light on the nature of the Jahn-Teller distortion, electronic states,
thermal behavior, and disorder in manganites \cite{Booth-PRB98,Qian-PRB00,Bridges-PRB01,Shibata-PRB03,Souza-PRB04,Sanchez-PRB06}.
The main parameters involved are the local coordination around Mn
atoms (Mn-O distances and distribution) and some slight modifications
on the octahedra arrangement. A consistent analysis of the nearest
neighbors does not necessarily require an a
priori knowledge of the long-range arrangement,
being otherwise able to provide masked details of the local structure.
We observe an hysteretic behavior that points to the coexistence of
two phases within a large pressure range (7 to 25~GPa). The ambient
phase with largely $JT$ distorted $MnO_{6}$ units coexist over a
large range of pressure with a new phase with more regular but still
distorted $MnO_{6}$ units. The IM transition takes
place within this latter distorted phase, without any significant
reduction of the local $JT$ distortion. Metallization turns out to
be unconnected with vanishing of the local distortion. Our study then
provides experimental evidences of a bandwidth-driven
nature of the pressure induced insulator to metal transition in LaMnO$_{3}$. 

High pressure XAS measurements at the Mn K-edge
(6539~eV) were performed at LUCIA beamline\cite{Flank-NIMB06} hosted
at the Swiss Light Source. The LaMnO$_{3.00}$ poly-crystal was synthesized
following an experimental procedure enabling an accurate control of
the oxygen stoichiometry \cite{Morales-JSSChem03}. The monochromatic
beam was focused to a 5x5$\mu m^{2}$ spot in a perforated diamond
anvil cell (DAC) with silicone oil as pressure-transmitting medium.
The non-hydrostatic components of this medium may lead to large pressure
deviations, up to 5\% over a diameter of 50 $\mu m$ for pressures
above 15 to 30 GPa \cite{shen-RSI2004,Klotz-JPD09}. However, it becomes
negligible in our experimental conditions where the highly focused
beam probes a small surface of 5x5$\mu m^{2}$. The pressure was \emph{in
situ} calibrated using the luminescence of a single
ruby chip nearly the DAC center, with an accuracy of \ensuremath{\approx}~0.3~GPa.
An error bar of 5\%, is estimated for the absolute pressure scale
due to the separation of the probed area and the ruby positions. Using
a gas membrane-driven mechanism, the pressure was increased step by
step up to 35~GPa, well above the expected IM transition, then released
by steps down to ambient pressure. EXAFS (Extended X-ray Absorption
Fine Structure) data were collected up to about 8~$\mathring{A}^{-1}$in
photo-electron wavenumber. The near-edge spectra (XANES) were normalized
between 150 and 250 eV above the edge.The sample
thickness was checked throughout the experiment and the normalization
corrects for a small sample thickness reduction with pressure. During
a sequence of XANES experiments, the energy shift was carefully monitored
by recurrent collections of the XANES spectrum of a Mn metal foil
(edge in 6539.1 eV). Moreover, the energy scale was verified after
releasing the pressure down to the ambient conditions; energy scale
stability as small as 0.1 eV were certified. The pre-edge and XANES
features were compared to \emph{ab initio} full multiple scattering
calculations using the FDMNES code\cite{Joly-PRB01} for 6.5\textrm{\AA}
large clusters using structural data of LaMnO$_{3}$ under ambient
conditions in the Pbnm structure\cite{Rodriguez-Carvajal-PRB98}.

\begin{figure}
\includegraphics[scale=1.0]{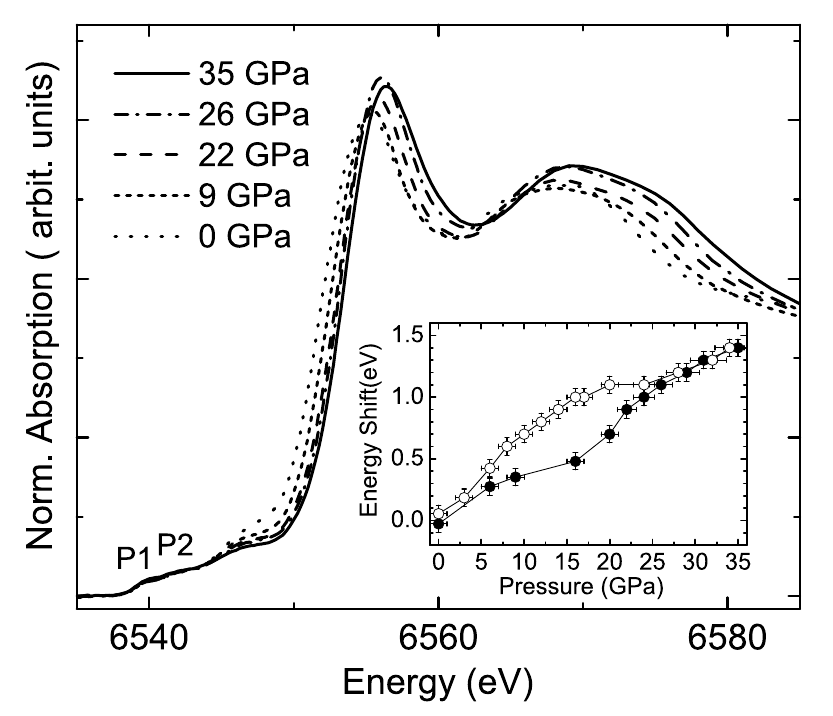}
\caption{Mn K edge XANES spectra for LaMnO$_{3}$
for selected pressures. The shift in the edge position is associated
to the reduction of the Mn-O long bond. The increase in the white
line (6555eV) corresponds to a decrease in JT splitting. Inset : edge
shift as a function of the applied pressure, for increasing pressure
(full symbols) and decreasing pressure (open symbols). }
\end{figure}

Figure 1 shows the Mn-K edge XANES at a few representative pressures.
The most striking features are the continuous shift of the absorption
threshold towards higher energies and the enhancement of the white
line (6555eV). Due to the selection rules in X-ray
absorption spectroscopy the 
edge transition originates from the core $1s$ state to the projected
$np$ (mainly $4p$) unoccupied density of states (DOS). XAS probes
the partial-DOS modified by the presence of the 1s core hole that
sorts out the 4p states around the Mn site from the ground state band
structure. The white line position and threshold depends then essentially
on the Mn-O bonds. The enhancement in the white line accounts for
the increasing overlap of the wave-functions imposed by hydrostatic
pressure. Such enhancement, well reproduced by \emph{ab initio} simulations,
indicates a reduction of the local distortion of the Mn sites \cite{Ramos-PRB07}.
These outcomes show qualitatively that the short bonds $(Mn-O){}_{s}$
are less reduced than the long $(Mn-O){}_{l}$ ones by the application
of an external pressure, in agreement with X-ray diffraction measurements\cite{Pinsard-PRB01,Loa-PRL01}.
As the manganese formal valence keeps unchanged, the edge shift ($\delta$E)
expresses modifications in the repulsive nearest neighbors potential
arising from shortening the $Mn-O$ bonds in the coordination shell\cite{Natoli-84}.
Besides, as shorter bond lengths correspond to higher edge energies,
the edge threshold is determined by the long bond
length $(Mn-O){}_{l}\approx$~2.15~$\textrm{\AA}$. Indeed,
the expected energy shift corresponding to the different long and
short distances ($\delta R\approx0.2\,\textrm{\AA}$) is about 2~eV\cite{SouzaNeto-PRB04}.
The evolution $\delta$E of the edge threshold as a function of the
applied pressure (inset fig. 1) is then essentially related to the
specific reduction in this long bond\cite{Ramos-PRB07}. The overall
shift of $\delta$E~=~+1.5~eV corresponds to a reduction of $\backsimeq$0.15~$\textrm{\AA}$
in bond length. The edge shifts strongly between 15 and 25~GPa when
pressure is increased but, when the pressure is released, a well-defined
hysteresis is observed (inset fig. 1). To recover the edge energy
of the initial point at 15~GPa, the pressure has to be released down
to about 5~GPa. Even if some deviatoric stress
effects might exist and lead to some strain in the small poly crystalline
grains\cite{shen-RSI2004,Klotz-JPD09}, these effects should be smaller
than the observed bond length hysteresis ($\thickapprox0.05\,\textrm{\AA}$
or 2.5\%) in the 10 to 15~GPa range. Zhao and coworkers \cite{Zhao-JAC2010}
found in a similar compound a highest strain difference of 0.3\% at
6 GPa originating from silicone oil at non-hydrostatic conditions.
XAS probes the local coordination around Mn atoms, and is not measuring
directly different phases. The hysteresis provides an indirect evidence
of the occurrence of mixed phases over a large pressure range. Such
hysteresis characterizes the first-order nature of a phase transition
where an activation barrier has to be overcome, causing the formation
of domains around nucleation centers. The spatial (domain size) and
temporal scales of these domains are probably too small to be observed
by diffraction techniques. Several papers in the literature relate
the presence of an hysteresis to the existence of phase mixing in
similar compounds \cite{Grunbaum-JSSChem04,SouzaJA-PRB04,Sanchez-PRB06,Chen-PRB10}.
The occurrence of mixed phases in $LaMnO_{3}$ under pressure has
already been pointed out by Baldini and coworkers\cite{Baldini-PhysRevLett2011}
in a similar pressure range. They identified peaks in the Raman spectra
related to phonon modes involving oxygen ions and a gradual transfer
of spectral weight from one phase to another.

\begin{figure}
\includegraphics[scale=1.0]{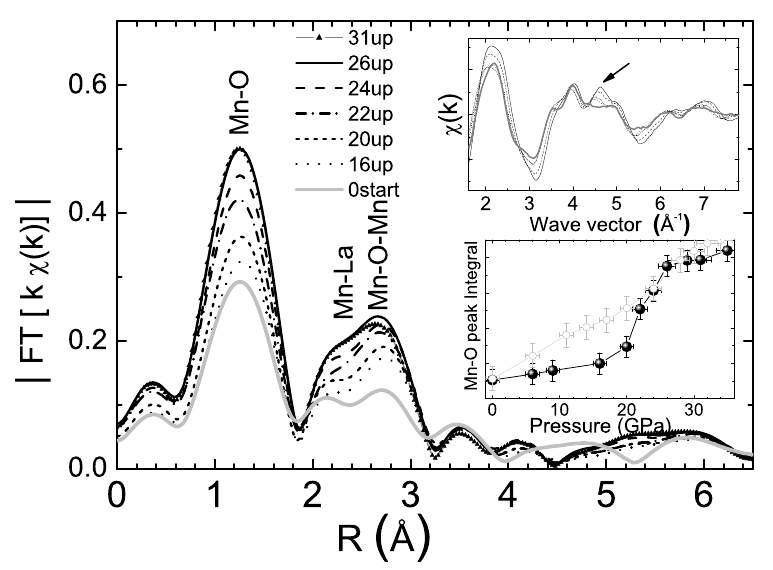}
\caption{Fourier Transform of the EXAFS signal for LaMnO$_{3}$
for selected pressures. Upper inset : EXAFS signal. Lower Inset :
evolution of the integral of the first peak, for increasing (plain
symbols) and decreasing (open symbols) pressures. }
\end{figure}

The hysteresis observed in the evolution of almost
all XANES features, is also observed in the EXAFS features. Figure
2 shows the Fourier Transform {(FT)}
of the experimental Mn-K edge EXAFS. The main peak contains information
on the Mn coordination shell, its position corresponds to the average
Mn-O bond length, not corrected for phase shifts. The amplitude of
the first peak directly reflects the disorder within the Mn coordination
shell. The restricted EXAFS k-range (8~$\mathring{A}^{-1}$) characterizes
a low-resolution study in R-space, where the Mn-O bond lengths are
no longer distinguishable and the bond length distribution appears
as a static contribution to the total disorder\cite{Araya-Rodriguez-JMMM01}.
The increase in this peak integral (fig.2, lower inset) corresponds
to a decrease in the separation between long and short Mn-O bond
lengths, i.e. in the $JT$ distortion. We identify three pressure
ranges: a first one of slow increase (P~<~15~GPa), a second one
of strong variation (15-25~GPa). Above 25 GPa the
first peak in the Fourier transform are almost superimposed ( FT at
26 GPa and 31GPa on figure 2), corresponding to a range of stabilization.
The large double peak at 2-3\textrm{\AA} contains essentially
the contributions of La single scattering and Mn-O-Mn multiple scattering.
The relative intensities in this double peak vary notably in the first
steps of the pressure increase (P < 7 GPa), then after only a global
increase is observed. This indicates that the various mechanisms of
inter-octahedra atomic accommodation (in particular octahedral tilts
and La shifts) take place mainly in the low pressure regime. This
is confirmed by the raise of the feature around $4.5\text{\AA}^{-1}$
in the EXAFS signal (fig.2, upper inset) associated to octahedral
tilt\cite{Souza-PRB04,Ramos-JPCS09} and agrees with previous X-ray
diffraction results\cite{Pinsard-PRB01}. We can deduce that, after
an initial octahedral rearrangement up to about 7~GPa, domains of
a new phase with less distorted octahedra start to form and coexist
with the initial phase up to 25~GPa, where only the second phase
persists.

\begin{figure}
\noindent \includegraphics[scale=1.0]{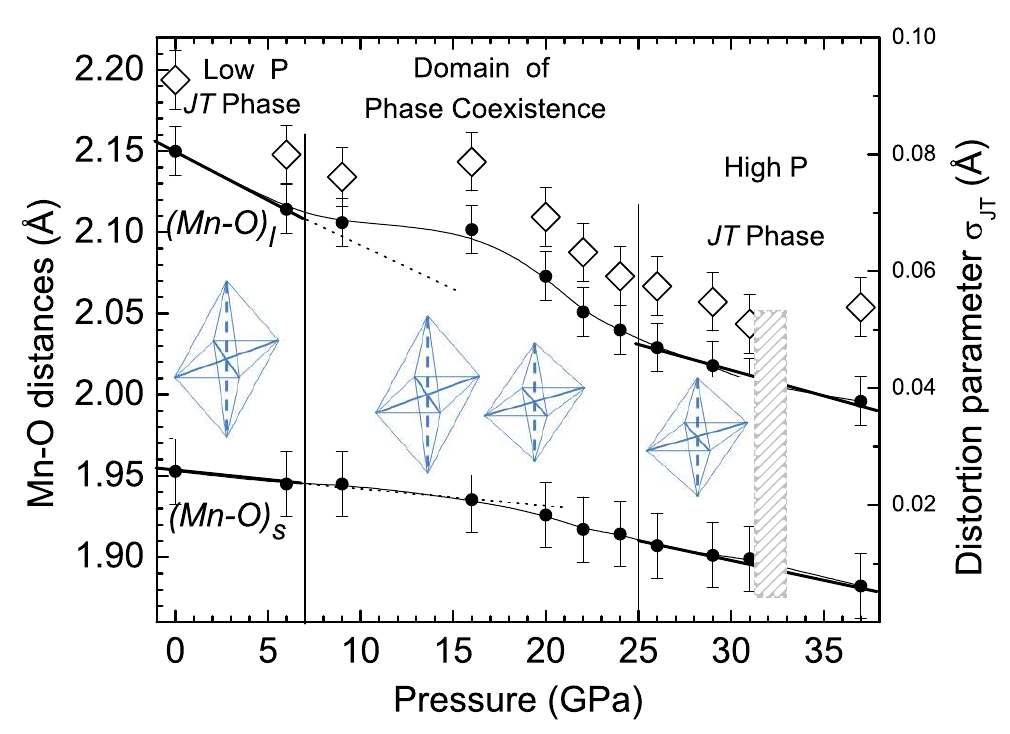}
\caption{Variation of the long $(Mn-O){}_{l}$ and short $(Mn-O){}_{s}$ bond
lengths (full circles) and distortion parameter
(open diamonds)with increasing applied pressure. The onset of the
metallic state is marked by an hatched zone. }
\end{figure}

To assess the parameters of these less-distorted $MnO_{6}$ units,
an overall quantitative analysis is done. For the
EXAFS signal there is an enormous weight on the shortest interatomic
distance occurring in the system, but no sensitivity to the long distance
tail\cite{Filipponi-JPC94}. In a bimodal distribution with 4 short
and 2 long distances, and given that the disorder associated to the
long bond is larger than the disorder associated to the short bond\cite{Souza-PRB04},
the two longer distances behave as a \textquotedblleft{}tail\textquotedblright{}
of the short bond signal. We checked the sensitivity of the EXAFS
analysis to such \textquotedblleft{} tail\textquotedblright{} by using
the software Feff7 \cite{Rehr-JAMChemSoc91} to simulate the EXAFS
signal given by Mn atoms in a distorted JT
environment. Due to limited available k range for the quantitative
EXAFS analysis the standard gaussian EXAFS fitting procedure on the
first peak of the Fourier Transform gives a single distance at the
value of the short one within an error of less than $0.02\,\textrm{\AA}$
, the longer distance contribution appearing as an slight increase
in the disorder parameter. On the other hand, in
the XANES range, the relationship between the reduction $\delta$R
of the long bond length and the associated edge shift $\delta$E is
rather linear \cite{SouzaNeto-PRB04} and allows a fair evaluation
of the long bond shortening with increasing pressure. From the edge
shift (1.5~eV) a maximum of $\approx0.15\,\textrm{\AA}$ is deduced.
Figure 3 reports the evolution of the long and short bond lengths
in the $MnO_{6}$ octahedra, obtained by this combined XANES and EXAFS
analysis. In the low pressure regime (P~<~7~GPa), the phase with
highly $JT$ distorted octahedra is present as a single phase. In
this phase the reduction of the bond length with pressure is anisotropic
and the $JT$ distortion rapidly decreases. This reproduces essentially
what has been reported by previous experiments\cite{Pinsard-PRB01,Loa-PRL01,Ramos-PRB07}.
The evolution of the short distance in the range 0 - 6 GPa is in good
agreement with the evolution of the average of the two shortest distance
found by Loa and coworkers\cite{Loa-PRL01}. For the long distance,
the agreement is also reasonable. The largest differences are found
above 7 GPa. The long bond found in the diffraction analysis keeps
dropping strongly, while in the XAS analysis, the decrease is retarded
in the domain where the hysteresis loop denounces a phase coexistence.
Taking as coherent distortion parameter $\sigma_{JT}$, defined as
: $\sigma_{JT}=\surd\frac{1}{6}\sum(Ri-R_{0})^{2}$, with $R_{i}$
individual distances and $R_{0}$ average distance within the coordination
shell, we found at ambient pressure $\sigma_{JT}$ \ensuremath{\approx}
0.1 and for P=35 GPa $\sigma_{JT}$ \ensuremath{\approx} 0.05 .
In between the evolution of the distortion parameter is given by the
open diamond shaped symbols in the figure 3. In the precedent XAS
study \cite{Ramos-PRB07} the data were limited in pressure and quality.
By extrapolating the evolution of the JT distortion up to 30 GPa we
guessed that the distortion could disappear. This extrapolation was
certainly spoiled by the presence of a phase mixing above 7 GPa. In
the present work the JT distortion parameter, that is an average parameter
over the phases presents in the compounds, is found to drop more slowly.
In the single phase present above 25~GPa, long and short Mn-O bond
lengths are still very different and their compression is nearly isotropic.
The bond length separation is almost unchanged with pressure. The
$MnO_{6}$ units present a lower $JT$ distortion. Pressure is less
effective in reducing the JT distortion above 25 GPa. In the range
0-5 GPa, associated to the distorted phase, $\sigma_{JT}$ drops by
about 15\% , i.e. a pressure dependence of $\delta\sigma_{JT}/\delta P\,\approx\,0.03$.
Over the range 26-35 GPa, associated to the second phase, it decreases
only by 0.005, i.e. $\delta\sigma_{JT}/\delta P\,\approx\,0.001$.
Especially, it keeps unchanged across the pressure range around 32~GPa,
where the IM transition is expected (fig.3, hatched area). 

We should mention that from EXAFS measurements we are not a priori  able to differentiate the small distortions due to JT  effect from site distortions intrinsic to a Pbnm space group. However the split preedge structure, discussed here below, is related to the lift of degeneracy of the 3d  $ e_{g}$ and $t_{2g}$  level by JT  effect, and the positions of the split components keep unchanged over the whole pressure range. We can then reasonably associate the small distortion at high pressure to a reduced JT  effect.

The present scenario is compatible with the recent
Raman data where the evidence of mixed phases comes from the emergence
of a new phonon peak and a gradual transfer of spectral weight from
the initial JT-peak to the new one. The strongest evolution in our
XAS data occurs within the 15 to 25~GPa range, which correlates to
the reversal of the spectral weight between Raman peaks.\cite{Baldini-PhysRevLett2011}
Nevertheless, we identify the new phase not as an undistorted but,
as a less-distorted $JT$ one. 

As far as theoretical approaches are concerned,
\textit{ab initio} calculations
using soft pseudo-potentials support the conservation of a local distortion
up to IM transition pressure \cite{Mizokawa-PRB99,Trimarchi-PRB05,Yamasaki-PRL06,Fuhr-PRL08}.
Trimarchi and coworkers predicted the occurrence of a structural phase
transition around 15~GPa leading to a situation with a mixture of
poly-types of antiferromagnetic
order \cite{Trimarchi-PRB05}. Yamasaki and co-workers \cite{Yamasaki-PRL06}
claimed by combining local density approximations (LDA) and mean field
theories that both on-site repulsion (d-d correlation) and JT
distortion are necessary for LaMnO$_{3}$ to be
insulating below 32~GPa. More recently, Fuhr and coworkers \cite{Fuhr-PRL08}
using LDA+U and a slave boson approach suggest close correlation between
the IM transition and the JT  local distortion;
however, they do not completely exclude the possibility of a JT
distorted metallic phase. These approaches have all been based on
experimental structural data at limited pressures\cite{Pinsard-PRB01,Loa-PRL01,Ramos-PRB07}.
We give here a firm basis for further theoretical studies. 

\begin{figure}
\noindent \includegraphics[scale=1.0]{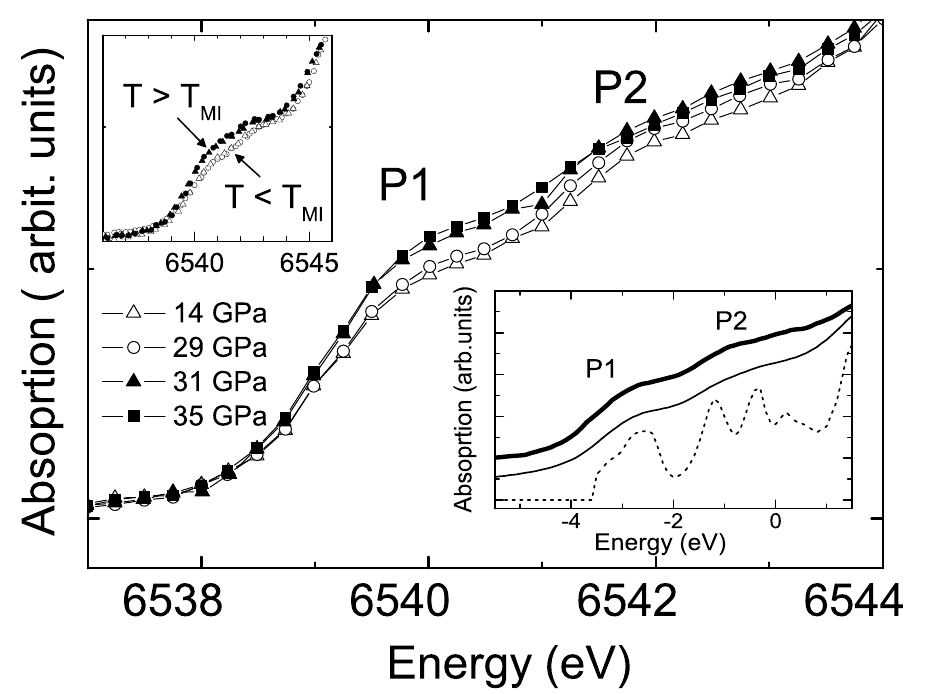}
\caption{Pre-edge features at increasing high pressures (open symbols P <
30 GPa, plain symbols P > 30~GPa). P1 and P2 peaks correspond respectively
to transitions towards unoccupied $e_{g}$ and $t_{2g}$ states. Upper
inset: pre-edge structure in LaMnO$_{3}$ around $T_{IM}$ (open symbols
: $T<T_{IM}$, plain symbols $T>T_{IM}$). Lower inset: comparison
between experience (thick line) and\emph{ ab initio} calculations
(thin line) of the pre-edge structures. The dotted line correspond
to the calculation before convolution. }
\end{figure}

We turn now to the pre-edge structures. In LaMnO$_{3}$, it consists
of two small peaks, assigned to dipole transition Mn $1s$ levels
to $4p$ empty levels\cite{Bridges-PRB00}~(figure 4). Ab initio
simulations confirm that the quadrupole contribution is insignificant.
$4p-3d$ hybridization for orbitals of the same Mn atom is forbidden
due to the centrosymmetry of the Mn sites. However the 4p orbitals,
having a large extension, hybridize with 3d orbitals of the Mn neighboring
atoms. The structures at the pre-edge reflect then the $3d$ partial
density of states. $P_{1}$ is associated to transitions to $e_{g}$
majority states, and $P_{2}$ to transitions towards $t_{2g}$ minority
states. Actually the $P_{2}$ peak contains two main contributions,
clearly evidenced in the derivative of the experimental spectrum (not
shown) and well reproduced in the calculated spectra (fig.4, lower
inset). Experimentally, we do not observe any change in the pre-edge
structure below 30~GPa. However, above 30~GPa, a small enhancement
in the first peak intensity is detected. Such enhancement is also
present across $T_{IM}$ for temperature-induced metallization\cite{Souza-PRB04}
(fig.4, upper inset). Similar enhancements of the $P_{1}$ peak have
also been reported in Ca-doped compounds \cite{Qian-PRB00,Ignatov-PRB01,Bridges-PRB00}.
It corresponds to an increase in the empty density of states associated
to metallization. $t_{2g}$ localized states being expected to participate
much less to the charge transfer, no change is detected in $P_{2}$.
One should also note that we do not observe any shift
in the first peak position, in contrast with the observations of Chen
and co-workers\cite{Chen-PRB09} in TbMnO$_{3}$. Such a shift, if
exists, should be lower than 0.1~eV. The enhancement of the $P_{1}$
structure provides an internal probe of the insulator to metal transition.
Therefore, we can clearly confirm that the IM transition takes place,
as expected, right above 30~GPa. It is then not directly associated
to any structural change, specially it is unconnected with the symmetrization
of the $MnO_{6}$ units. 

In summary, we have investigated the local order in LaMnO$_{3}$ under applied pressure. 
The ambient phase with largely
$JT$ distorted $MnO_{6}$ units coexist over a large pressure range
with a new phase with more regular but still $JT$ distorted $MnO_{6}$
units. This latter phase is present alone above 25~GPa. No structural
anomaly or singularity is observed in this pressure range. On the
other hand, the IM transition around 32~GPa is finger-printed from
the XAS pre-edge structures. Our study provides evidences that the
$IM$ transition takes place in a distorted $JT$ phase, without quenching
of this distortion across the transition, which characterizes a unconventional
Mott insulator. Insulator state and local distortion are not totally
correlated and metallization should necessarily be driven by band
overlap imposed by decreasing bond lengths under pressure. 

We acknowledge financial support by the European Community for the
experiments at SLS (Swiss Ligth Source).

\end{document}